\documentclass[fleqn,12pt,twoside]{article}
\usepackage{espcrc1}

\usepackage{graphicx}
\usepackage[figuresright]{rotating}

\newcommand{\AmS}{{\protect\the\textfont2
  A\kern-.1667em\lower.5ex\hbox{M}\kern-.125emS}}

\title{Nucleosynthesis as a result of multiple delayed detonations in  
       Type Ia Supernovae}

\author{Domingo Garc\'\i a-Senz
 \address{Dpt. de F\'\i sica i Enginyeria Nuclear. UPC. Jordi Girona 1-3,\\  
      M\`odul B5, 08034 Barcelona, Spain
            }\address{Institut d'Estudis Espacials de Catalunya, Gran Capit\`a 
	       2-4, 08034 Barcelona}
            and
            Eduardo Bravo$^{\rm {\ a b}}$}

\begin{document}

\maketitle

\begin{abstract}
The explosion of a white dwarf of mass 1.36 M$_\odot$~has been simulated in 
three dimensions with the aid of a SPH code. The explosion follows the delayed 
detonation paradigma. In this case the deflagration-detonation transition is 
induced by the large corrugation of the flame front resulting from 
Rayleigh-Taylor instability and turbulence. The nucleosynthetic yields have 
been calculated, 
showing that some neutronized isotopes such as $^{54}Fe$~ or $^{58}Ni$~ are   
not overproduced with respect to the solar system ratios. The distribution of 
intermediate-mass elements is also compatible with the spectra of normal SNIa. 
The excepcion is, however, the abundance of carbon and oxygen, which are 
overproduced.

\end{abstract}

\section{Introduction}

The undeniable influence of Type Ia Supernovae explosions in, among others,
the chemical
evolution of the galaxy and cosmology makes the quest of solid
theoretical models as an urgent task. Guided by the insight gained from 
two decades of modelization of spherically symmetric models, modern
multidimensional calculations try to make a reliable picture of the explosion
on physical basis, minimizing the number of free parameters. Nevertheless, even 
supossing that the exploding object is a CO white dwarf near to the
Chandrasekhar-mass limit the subject is not firmly resolved yet. In particular, 
there remain uncertainties in the immediate pre-explosive stage and in the 
mechanism/s by which the thermonuclear flame is spreaded through the white 
dwarf. In this regard a difficult point is related with the acceleration of 
the combustion. We know that the explosion is triggered by a thermonuclear
flame whose initial velocity is clearly subsonic. We also know, from
parametrized calculations, that after several sound-crossing times the
effective velocity of the deflagration must reach a significant fraction of 
of the local sound speed in order to synthesize the 
adequate amount of nuclear species to explain both light curve and spectra.
This translates into an increase factor of thousands over the pure laminar 
value at maximum effective velocity. The physical mechanism behind this
huge factor is 
currently attributed to flame surface corrugation by hydrodynamical
instabilities (mainly Rayleigh-Taylor and turbulence). A branch in this 
scenario could be achieved if the already fast deflagration turns into a 
steady detonation at late times, when flame density is below 
$\rho\simeq 4~10^7$~g.cm$^{-3}$. This is the so called delayed-detonation 
scenario which, in calculations carried out in spherical symmetry, have
provided some of the best
models for SNIa in the past \cite{hk96}. 

In this short paper we present the results of a calculation concerning the 
delayed detonation model carried 
out in three dimensions by using a SPH code. We describe the main features of 
the evolution of the model with more emphasis in the nucleoshynthetical side.

 \begin{table*}
\hfill \parbox[b]{5cm}{}
\caption[]{Main features of the model$^{\mathrm{a}}$
} \label{Table 1}
\smallskip
\begin{tabular}{crrrrrrr}
\hline
\noalign{\smallskip}
t(s) &$\rho_c$~(g.cm$^{-3})$&$\left<\rho_{\mathrm flame}\right>$~(g.cm$^{-3})$&
$E_{\mathrm kin}$~$(10^{51}\mathrm erg)$&$M_{\mathrm CO}(M_\odot)$&
$M_{\mathrm Si}(M_\odot)$&
$M_{\mathrm Ni} (M_\odot)$ \\
\noalign{\smallskip}
\hline
\noalign{\smallskip}
0&$1.4~10^9\quad$&$1.06~10^9\qquad$&$2.6~10^{-3}\qquad$&$1.29\qquad$&$3.4~10^{-4}$&
$1.84~10^{-3}$\\
1.54&$5.4~10^7\quad$&$2.04~10^7\qquad$&$0.29\qquad\quad$&$0.65\qquad$&
$0.072\quad$&$0.272\quad~$\\
7.92&$3.5~10^3\quad$&$-\qquad\qquad$&$0.75\qquad\quad$&$0.34\qquad$&
$0.16\quad~$&
$0.54\qquad$\\
\noalign{\smallskip}
\hline
\end{tabular}
\begin{list}{}{}
\item[$^{\mathrm{a}}$]
First column is the elapsed time since the beginning of the SPH simulation.    
Previously, the initial stages of the explosion were  calculated with
a Lagrangian one-dimensional code.
\end{list}
\end{table*}

\section{Description of the model}

The initial model is a white dwarf of $1.36$~M$_\odot$~in hydrostatic
equilibrium. The first stages of the explosion were followed by using a 
1D hydrocode until the central density declined to $\rho_c=4~10^7$~g.cm${-3}$. 
From here on the model was mapped into a 3D distribution of 250,000 particules 
and its evolution followed with the SPH. In addition, the velocity field 
around the flame front was perturbed at the beginning in order to seed
the hydrodynamical 
instabilities. This allows us to describe the large scale deformation of the 
flame in a self-consistent way although a lot of surface is lost owing to the
limited resolution of the code. To take into account 
these hidden lengthscales a subgrid model has to be set. We 
assume that the main effect of the smallest lengthscales is to increase   
the laminar velocity of the flame, $v_{\mathrm l}$, until a first
effective value
$v_{\mathrm b}>v_{\mathrm l}$. Afterwards the final effective velocity
$v_{\mathrm eff}>v_{\mathrm b}$~is directly computed from the hydrocode 
\cite{gb03}. 

During the progression of the calculation the geometrical features of the 
flame front were tracked by calculating the main scaling parameters of the 
surface such as $l_{\mathrm min}$, $l_{\mathrm max}$~and flame fractal 
dimension $D$~ by using the method described in \cite{gbs98}. As discussed in 
\cite{gb03} the 
local evolution of $D$~can be used as a practical criteria to turn the 
deflagration into a detonation. In particular, we have allowed such  
transition in those regions of the flame with fractal dimension higher
than 2.5 once flame mean density is $\left<\rho_{\mathrm flame}\right>\simeq 2~10^7$~g.cm$^{-3}$. 

The relevant features of the evolution of the model are summarized in Table 1. 
Basically the explosion proceeded in two phases. From t=0 s to t=1.54 s the 
combustion propagated subsonically as a deflagration. However, a considerable
acceleration of the flame was seen because the effective flame velocity passed 
from $v_{\mathrm eff}\simeq 8~10^{-3} c_s$~ for $t<0.7 s$~ to
$v_{\mathrm eff}\simeq 0.14 c_s$~at
$t=1.41 s$. At the end of this phase the total energy was already positive,  
although too low to represent a normal supernovae explosion. At t=1.54 s a 
detonation was artificially induced in those regions of the flame surface 
which had a large fractal dimension. In consequence multiple detonations 
emerged from unconnected points spreaded on a very irregular surface.  

Information about the nucleosynthetic yields is provided in Table 1 and 
figures 1,2. The ejected amount of $^{56}\mathrm Ni$~is enough to power 
the light curve of a typical Type Ia supernovae and the amount of silicon and 
other intermediate-mass elements is consistent with the spectrum near maximum
light. 
A clear difference with respect 1D calculations is that the  abundance 
distribution in velocity space show a lot of dispersion in 
comparison. The isotopic composition of the 
ejecta, calculated by postprocessing the dynamical model, is depicted in 
figure 2. Although we see a large production of vanadium there is not
overproduction of problematic isotopes such as
$^{54}Fe$~ or $^{58}Ni$. This is a feature of multidimensional models where 
electron captures proceed,on average, at lower density owing to the buoyancy 
of high entropy ashes. The 
ejected amount of carbon and oxygen is somewhat high, being distributed in 
isolated pockets within the star but not close to the center. Our results  
suggest that the three-dimensional version of the delayed detonation model
is also able to reproduce the main 
observational features of Type Ia Supernovae.  

This work has been benefited from the MCYT grants EPS98-1348 and AYA2000-1785 
and by the DGES grant PB98-1183-C03-02.

\begin{figure}
\includegraphics[width=20pc]{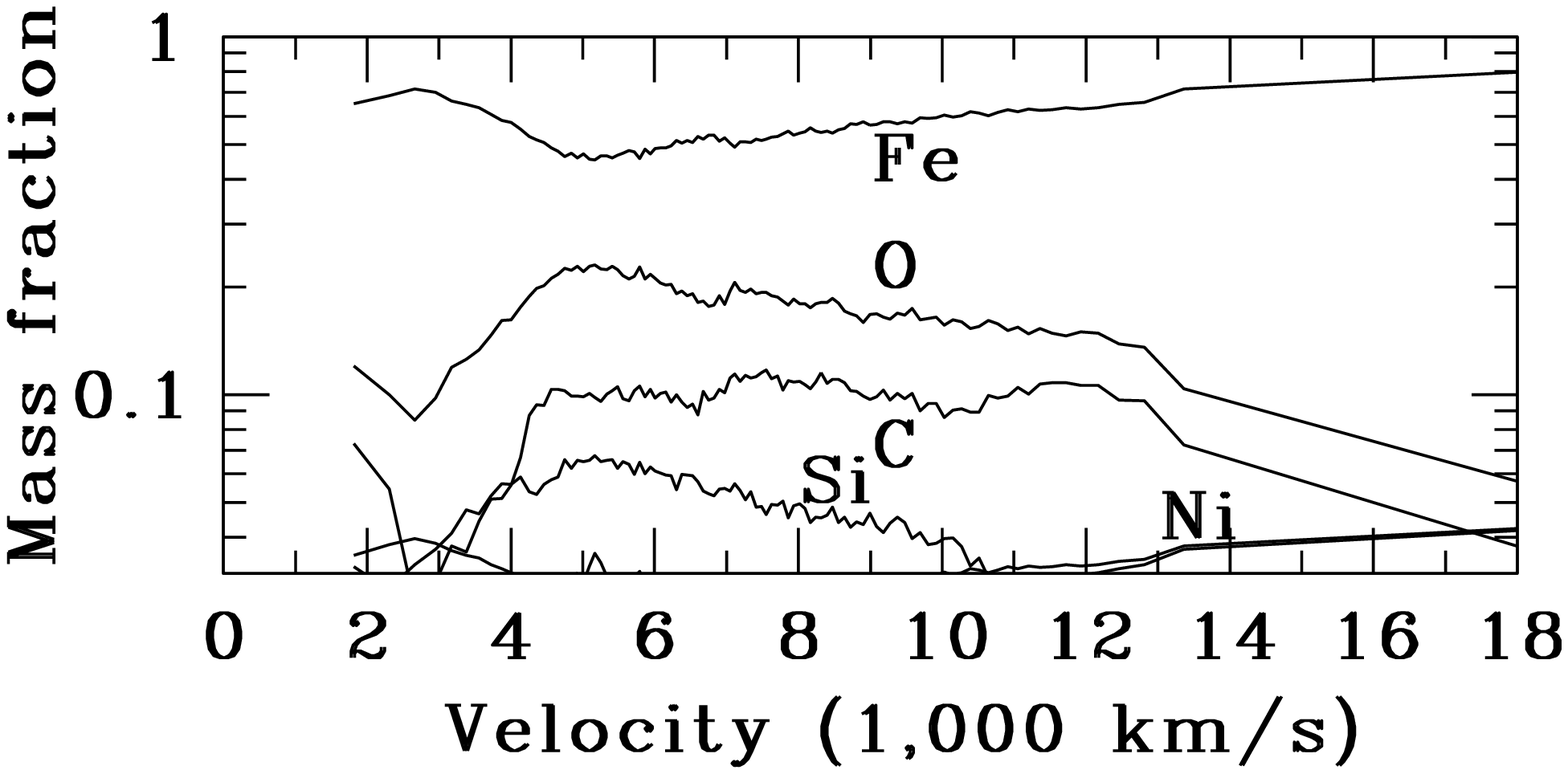}
\includegraphics[width=20pc]{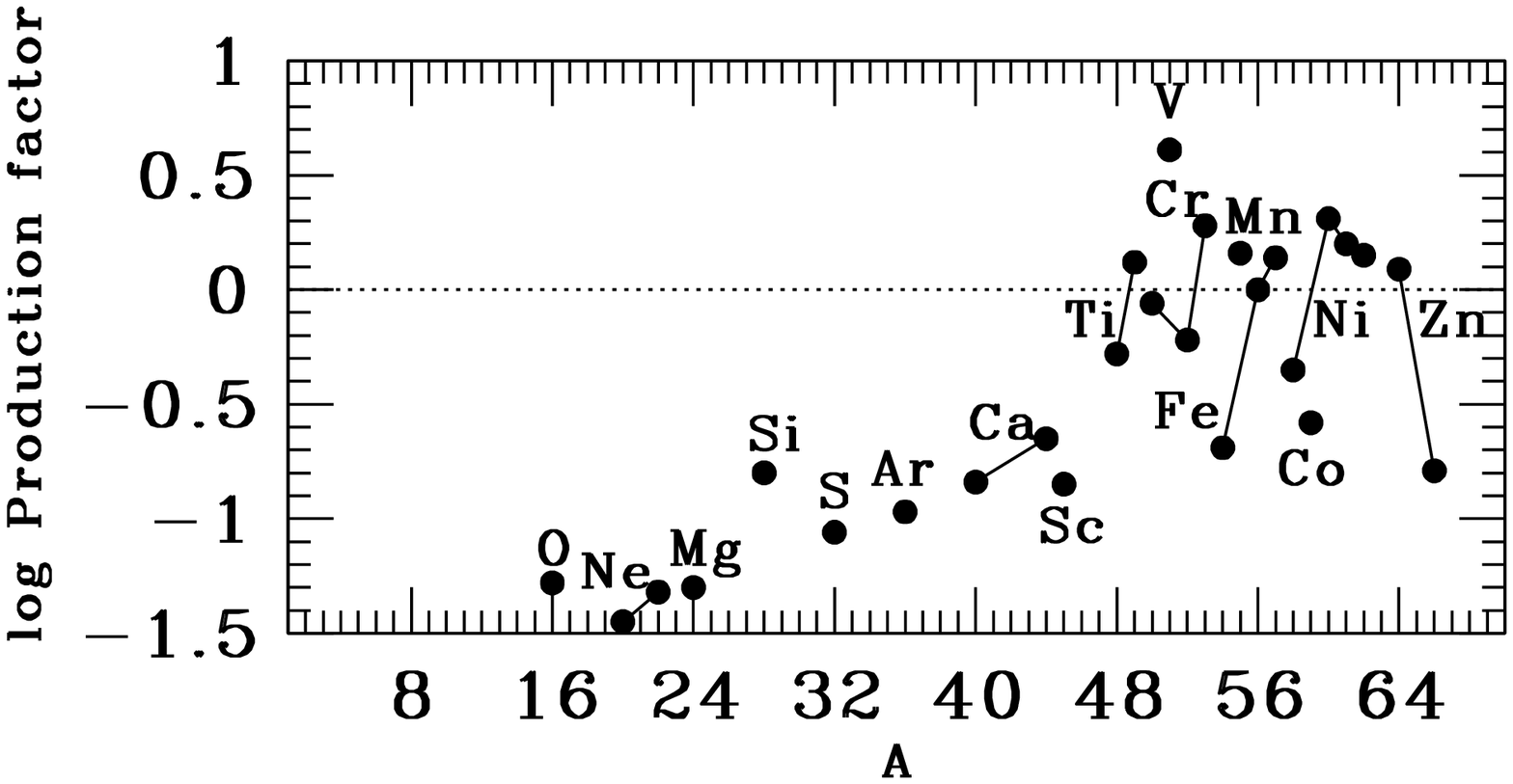}
\begin{minipage}[t]{75mm}
\caption{Distribution of nuclear species in velocity space} 
\end{minipage}
\hspace{\fill}
\begin{minipage}[t]{75mm}
\caption{Final production factors of several isotopes normalized to solar 
system abundances
}
\end{minipage}
\end{figure}


\begin{thebibliography}{9}
\bibitem{hk96} P. H\"oflich and A. Khokhlov, ApJ 457 (1996) 500.
\bibitem{gb03} D. Garc\'\i a-Senz and E. Bravo, in Proceedings of 
		the ESO/MPA/MPE workshop "From Twilight to Highlight: The 
		 Physics of Supernovae", (2003), in Press.
\bibitem{gbs98} D. Garc\'\i a-Senz, E. Bravo and N. Serichol, ApJSS 115 (1998) 
		119
\end{thebibliography}
\end{document}